\documentstyle[11pt,epsf,graphicx]{article}
\def\r{\mbox{\boldmath $r$}}

\def\p{\mbox{\boldmath $p$}}
\def\q{\mbox{\boldmath $q$}}

\begin{document}

\begin{center}
{\Large\bf{Recent Advances in the Description of Electromagnetic Two-Nucleon 
Knockout Reactions}
\footnote[1]{presented by C. Giusti, E-mail: giusti@pv.infn.it, 
phone: +39 0382987454, fax: +39 0382526938.}}
\end {center}
\vskip0.5truecm
\centerline{\large{C. Giusti,  F.D. Pacati}}
\vskip0.5truecm
\centerline{\small{Dipartimento di Fisica Nucleare e Teorica, 
Universit\`a degli Studi di Pavia,}} 
\centerline{\small{and Istituto Nazionale di Fisica Nucleare, Sezione di Pavia, 
I-27100 Pavia, Italy}} 
\vskip0.5truecm
\centerline{\large{M. Schwamb}}
\vskip0.5truecm
\centerline{\small{Dipartimento di Fisica, Universit\`{a} degli Studi di 
Trento,}} 
\centerline{\small{and Istituto Nazionale di Fisica Nucleare,
Gruppo Collegato di Trento,  I-38100 Povo (Trento),}} 
\centerline{\small{and European Center for Theoretical Studies in Nuclear
Physics and Related Areas ($\mathrm{ECT^*}$)}}   
\centerline{\small{ I-38100 Villazzano (Trento), Italy}} 

\begin{abstract}
Recent advances in the description of electromagnetic two-nucleon knockout 
reactions are reviewed. The sensitivity to different types of correlations and 
to their treatment in the nuclear wave functions, the effects of final-state 
interactions and the role of center-of-mass effects in connection with the 
problem of the lack of orthogonality between initial bound states and final 
scattering states obtained by the use of an energy-dependent optical-model 
potential are discussed. Results are presented for proton-proton and
proton-neutron knockout off $^{16}$O also in comparison with the available 
data. 
\end {abstract}
\vspace{1cm}

\section{Introduction}

Since a long time electromagnetic two-nucleon knockout reactions have been
devised as a preferential tool to investigate two-body correlations in nuclei 
\cite{Gott,Oxford}. 
Indeed, the probability that a real or virtual photon is absorbed by a pair of
nucleons should be a direct measurement of the correlations between the two
nucleons.
For an exclusive reaction the transitions amplitudes contain 
the two-nucleon overlap function (TOF) between the ground state of the target and the 
final state of the residual nucleus.  In the cross section the TOF gives the two-hole 
spectral density function, {\it i.e.}
\begin{equation}
S(\bar{\p}_1,\bar{\p}_2\p_1,\p_2,;E_{\mathrm m})=
\langle \Psi_{\mathrm{i}} |a^+_{\p_2} a^+_{\p_1}
\delta(E_{\mathrm m} -H)a_{\bar{\p}_1} a_{\bar{\p}_2}| \Psi_{\mathrm{i}} \rangle,
\label{eq:H2SF}
\end{equation}
which in its diagonal form ($\p_1=\bar{\p}_1$ and $\p_2=\bar{\p}_2$)
gives the joint probability of removing from the target two nucleons, with
momenta $\p_1$ and $\p_2$, leaving the residual nucleus in a state with energy 
$E_{\mathrm m}$ with respect to the target ground state. 
In an inclusive reaction, integrating the spectral density function over the 
whole energy spectrum produces the two-body density matrix 
$\rho(\bar{\p}_1,\bar{\p}_2,\p_1,\p_2)$, that in its diagonal form
and in the coordinate representation gives the pair correlation function 
$f(\r_1,\r_2)$, {\it i.e.} the conditional probability density of finding in the 
target a particle at $\r_2$ if another one is known to be at $\r_1$.

The two-nucleon overlap, the two-hole spectral function and the two-body density
matrix contain information on nuclear structure and correlations.  
The problem is that it is not easy to disentangle these quantities from the 
experimental cross sections and extract clear and unambiguous information on
correlations. In order to achieve this goal, along with the experimental 
work, a reliable theoretical model is needed, where all the ingredients are well under control. 

By correlations we mean what goes beyond the independent-particle model, the
mean-field approximation. Of particular interest are the short-range
correlations (SRC), which are produced by the short-range repulsion of the
NN-interaction. SRC are of particular relevance for pp-pairs and
are essential for understanding the short-range and high-momentum properties of
the nuclear wave functions. Also very interesting are the tensor correlations
(TC), which are due to the strong tensor component of the NN-interaction, 
that is of intermediate to long-range character. TC are of particular relevance
for pn-pairs. Thus SRC and TC can be investigated for pp- and pn-pairs. Such an
investigation would be of great interest, since the short-range repulsion and 
the tensor character represent the two main features of the NN-interaction. 
Also very important are, however, the so-called long-range correlations (LRC) 
at low energy, which mainly represent the interaction of nucleons at the nuclear 
surface and are related to the coupling between the single-particle (s.p.) 
dynamics and the collective excitation modes of the nucleus. SRC, TC, and LRC 
are intertwined in the spectral function and in the density matrix and it is 
not easy to disentangle a specific contribution. A reliable  
model for cross section calculations requires a careful and consistent 
treatment of these different types of correlations. Moreover, the cross section
is sensitive to the reaction mechanism.

There are two competing processes contributing to two-nucleon knockout.
There is the process where the electromagnetic probe hits, through a one-body
(OB) current, either nucleon of a correlated pair and both nucleons are then 
ejected from the nucleus. Two nucleons cannot be emitted by an OB current unless
they are correlated. Thus, this process is entirely due to correlations.  
Two nucleons, however, can also and naturally be emitted by two-body (TB) 
currents, due to meson exchanges and $\Delta$-isobar excitations, even if they
are not correlated. The role and relevance of these two competing processes can
be different in different situations and kinematics, but if situations can be
envisaged where the OB current is 
dominant, these situations are very well suited to investigate correlations. 

It must be noted, however, that correlations affect also the process due to the
TB currents. In the model \cite{GP,GP97,Giu97,barb} the transition amplitudes
contain three main ingredients: the nuclear current, the two-nucleon scattering
wave function, and the TOF, which includes correlations. The two-nucleon wave 
functions for the initial and final states are consistently treated in the 
transition amplitudes with the OB and the TB currents. Thus, correlations 
affect also the contribution of the TB currents and in principle it is possible 
to obtain information on correlations even when the TB currents are important or 
dominant in the cross section. 

Moreover, it is not correct to say that the contribution of the OB current is 
entirely due to correlations in the nuclear wave functions. In the calculation 
of the transition amplitudes, for the three-body system consisting of the two 
nucleons and the residual nucleus, it is natural to work with center-of-mass 
(CM) coordinates, and in the CM frame a TB operator is obtained in the 
transition  amplitude even with an OB current \cite{cm1,cm2}. The relevance of 
this CM effect depends on the particular situation that is considered and
may be large or small, but, independently of that, due to this CM effects 
there is a contribution to two nucleons emission with an OB current also 
without correlations in the nuclear wave functions.     

The main features of the theoretical approach  are outlined in Sec. 2. 
Special emphasis is devoted to the description of the recent improvements 
performed in comparison with earlier work. Numerical results for the exclusive 
pp- and pn-knockout reactions off $^{16}$O are presented and discussed in 
Sec. 3 also in comparison with the available data. Some conclusions are drawn 
in Sec. 4.

\section{Theoretical Framework}
\label{sec:frame}

The basic ingredients for calculations of the reaction induced by a real or 
virtual photon, with momentum \q, where two nucleons, with momenta
$\p^{\prime}_1$ and  $\p^{\prime}_2$, are emitted from a nucleus, 
are the transition matrix elements of the charge-current density 
operator between initial and final nuclear states, {\it i.e.}
\begin{equation}
J^{\mu}(\q) = \int \left\langle \Psi_{\mathrm{f}} \right| \hat{J}^{\mu}(\r)
\left| \Psi_{\mathrm{i}} \right\rangle {\mathrm{e}}^{\,{\mathrm{i}} 
{\footnotesize {\q}} \cdot {\footnotesize {\r}}} {\mathrm d}\r  .
 \label{eq:jm}
\end{equation}
Bilinear products of these integrals give the components of the hadron tensor
whose suitable combinations give all the observables available from the
reaction process.

The model is based on the two assumptions of an exclusive reaction, where the 
residual nucleus is left in a discrete eigenstate of its Hamiltonian, and of 
the direct knockout mechanism. As a results of these two assumptions, the 
matrix elements of Eq.~(\ref{eq:jm}) are obtained in the form  \cite{GP,GP97}
\begin{equation}
J^{\mu}(\q) = \int \psi_{\mathrm{f}}^{*}(\r_1,\r_2) J^{\mu}(\r,\r_1,\r_2)
\psi_{\mathrm{i}}(\r_1,\r_2){\mathrm{e}}^{\,{\mathrm{i}} {\footnotesize {\q}} 
\cdot {\footnotesize {\r}}} {\mathrm d}\r {\mathrm d}\r_1 {\mathrm d}\r_2 .
 \label{eq:jq}
\end{equation}
Equation~(\ref{eq:jq}) contains three main ingredients: the nuclear current 
$J^{\mu}$, the two-nucleon scattering wave function $\psi_{\rm{f}}$, and the 
TOF $\psi_{\rm{i}}$. In the model 
$\psi_{\rm{i}}$ and $\psi_{\rm{f}}$ are consistently derived from an 
energy-dependent non-Hermitian Feshbach-type Hamiltonian  \cite{Oxford,GP}. 
Actually, they are eigenfunctions of this Hamiltonian and of its Hermitian 
conjugate at different energies: for an exclusive reaction, we select a 
channel and then project the initial and final nuclear states in 
Eq.~(\ref{eq:jm}) onto the selected channel subspace; as a result of the 
projection, we obtain the optical-model Hamiltonian describing the interaction 
of the two nucleons and the residual nucleus in the eigenstate that we have 
selected \cite{Oxford,GP}. 
Since a fully consistent calculation of $\psi_{\rm{f}}$ and $\psi_{\rm{i}}$  as
eigenfunctions of the optical-model Hamiltonian would be very difficult, 
some approximations are used in actual calculations. 

\subsection{Nuclear Current}
\label{sec:Current}

The nuclear current operator is the sum of an OB and a TB part. The OB part 
contains the longitudinal charge term and the convection and spin 
currents. The TB currents are derived from a nonrelativistic reduction of the 
lowest-order Feynman diagrams with one-pion exchange. Therefore we have 
currents corresponding to the seagull and pion-in-flight diagrams, and to the 
diagrams with intermediate $\Delta$-isobar configurations \cite{Giu98}. 
All these terms contribute to pn-knockout, while the seagull and the 
pion-in-flight meson-exchange currents (MEC) do not contribute to pp-emission, 
at least in the adopted nonrelativistic limit.

The contribution of the TB currents depends on the type of reaction, on the
kinematics, on the conditions of the calculations, and, of course, on the
treatment of the TB currents. The explicit expressions of the currents can be
found in \cite{Giu98,WiA97,delta}. 
Different parametrizations have been used in previous 
calculations, in particular for the  $\Delta$-current. A fundamental treatment 
of the $\Delta$ in electromagnetic breakup reactions on complex nuclei is 
presently unavailable and we can only rely on approximative schemes.  

In our most recent calculations \cite{cm1,cm2,delta}, the parameters of the 
$\Delta$-current are fixed considering the NN-scattering in the 
$\Delta$-region, where a reasonable  description of scattering data is achieved 
with parameters similar to the ones of the full  Bonn potential \cite{MaH87}.
For pp-knockout, besides the usual $\pi$-exchange, also $\rho$-exchange 
is  included and the coupling constants are
\begin{equation}
\frac{f^2_{\pi \mathrm{NN}}}{ 4\pi} =0.078, \,\,\, 
\frac{f^2_{\pi \mathrm{N} \Delta}}{ 4\pi} =0.224, \,\,\,
\frac{f^2_{\rho \mathrm{NN}}}{ 4\pi} =7.10, \,\,\,
\frac{f^2_{\rho \mathrm{N}\Delta}}{ 4\pi} =20.45. 
\label{parapp}
\end{equation}
Moreover, hadronic form factors are included 
\begin{equation}
 F_{x {\mathrm{NN}}}(q^2) = \left(\frac{ \Lambda^2_{x {\mathrm{NN}}} - 
 m^2_{x}}{\Lambda_{x{\mathrm{NN}}}^2 +
 q^2 }\right)^{n_{x {\mathrm{NN}}}}, \,\,\,  F_{x \mathrm{N} \Delta}(q^2)  = 
 \left(\frac{ \Lambda^2_{x \mathrm{N} \Delta} - m^2_{x}}
 {\Lambda_{x \mathrm{N}\Delta}^2 +
 q^2 }\right)^{n_{x \mathrm{N}\Delta}},
 \label{form}
\end{equation}
with $n_{x {\mathrm{NN}}} = n_{x \mathrm{N}\Delta}= 1$, and the cutoffs   
\begin{equation}
\Lambda_{\pi \mathrm{NN}} = 1300 \,\mbox{MeV}, \,\,\, 
\Lambda_{\pi \mathrm{N} \Delta} = 1200 \,\mbox{MeV}, \,\,\,
\Lambda_{\rho \mathrm{NN}} = 1400 \, \mbox{MeV},  \,\,\,
\Lambda_{\rho \mathrm{N} \Delta} = \, 1000\, \mbox{MeV}.
\label{formpp}
\end{equation}

For pn-knockout, where the TB current contains many more terms, a
simpler approach is adopted \cite{cm2} where only $\pi$-exchange is included
 and the parameters for the $\Delta$-current are 
\begin{eqnarray}
\frac{f^2_{\pi \mathrm{NN}}}{ 4\pi} =0.078, \, & &
 \frac{f^2_{\pi \mathrm{N} \Delta}}{ 4\pi} =0.35, \nonumber \\
 n_{\pi \mathrm{NN}}=n_{\pi \mathrm{N} \Delta}=1, \, & &  
 \Lambda_{\pi \mathrm{NN}}= \Lambda_{\pi \mathrm{N} \Delta}=700 \, \mbox{MeV}. 
\label{formpn}
\end{eqnarray}
These parameters are able to give, with only $\pi$-exchange, a 
comparable description of the NN-scattering data.   
With respect to the pion-in-flight and seagull MEC, contributing
 only to pn-knockout, a dipole cutoff  is used of 3 GeV, in accordance 
 with the Bonn-C potential \cite{MaH87} which is used also in  
the calculation of the TOF \cite{barb} and of the mutual interaction 
between the two outgoing nucleons \cite{sch1,sch2}.
This choice of parameters for the TB currents is called in the following 
Bonn parametrization. 
 
The hadronic form factors defined in Eq.~(\ref{form}) are necessary for 
regularizing the interaction at short distances, where the meson-exchange 
picture becomes meaningless.
In previous calculations \cite{Giu97,barb,Giu99} the same coupling 
constants as in Eq.~(\ref{formpn}) were used with a simpler regularization in 
coordinate space, both for the MEC and the $\Delta$-current, which in practice 
is similar to an unregularized prescription 
($\Lambda_{\pi \mathrm{NN}} = \Lambda_{\pi \mathrm{N} \Delta} \rightarrow
\infty$).

\subsection{Final-State Interaction}
\label{sec:FSI}

In  the scattering state $\psi_{\rm{f}}$ the two outgoing nucleons, 1 and 2, 
and the residual nucleus interact via the potential
\begin{equation}
V_{\mathrm{f}}=V^{\mathrm{OP}}(1) + V^{\mathrm{OP}}(2) + V^{\mathrm{NN}}(1,2),
\label{eq:pot}
\end{equation}
where $V^{\mathrm{OP}}(i)$ denotes the interaction between the nucleon $i$ and 
the residual nucleus, described in the model by a complex optical potential, 
and  $V^{\mathrm{NN}}(1,2)$ denotes the mutual interaction between the two 
outgoing nucleons. Only the contribution of the final-state interaction (FSI) 
due to $V^{\mathrm{OP}}(i)$ was included in our first calculations. In this 
approximation (DW) the two-nucleon scattering wave function is given by the 
product of two uncoupled s.p. distorted wave functions, eigenfunctions of a 
phenomenological optical potential fitted to nucleon-nucleus scattering data 
\cite{Nad}. The contribution of  $V^{\mathrm{NN}}(1,2)$ (NN-FSI) has been 
studied within a perturbative approach  \cite{sch1,sch2}. 

The main contribution of FSI is given in general by the optical potential, 
which produces an overall and substantial reduction of the calculated cross 
sections. This effect is important and can never be neglected. 
NN-FSI gives in general an enhancement of the cross section that depends 
strongly on the kinematics, on the type of reaction, and on the final state of 
the residual nucleus. It is generally larger in pp- than in pn-knockout and in 
electro- than in photoinduced reactions \cite{sch1,sch2}. 
In many situations the difference between the results obtained in the DW 
approach and in the more complete approach where also NN-FSI is included 
(DW-NN) is small. There are, however, also situations where this difference 
is large. Numerical examples are presented in Sec. 3.

\subsection{Two-Nucleon Overlap Function}
\label{sec:TOF}

The TOF $\psi_{\rm{i}}$ contains information on nuclear structure and
correlations and represents the most interesting ingredient of the model. 
Different approaches have been used  
\cite{GP97,Giu97,barb,Giu98,
Giu99,Kadrev} to calculate $\psi_{\rm{i}}$  in pp- 
and pn-knockout from $^{16}$O. 
$^{16}$O is a suitable target for this study, due to the presence 
of discrete final states in the excitation-energy spectrum of the residual 
nucleus, both 
for $^{14}$C and $^{14}$N, states well separated in energy and that can be 
separated in experiments with good energy resolution.
Experimental data are available for both $^{16}$O(e,e$'$pp) 
\cite{Ond,Sta00,Ros00} and $^{16}$O (e,e$'$pn) \cite{Duncan} reactions.

In a simpler approach \cite{GP97,Giu98} the TOF is given by the product of a 
coupled and antisymmetrized shell-model (SM) pair function and  
a Jastrow-type central and state independent correlation function taken 
from \cite{GD}. In this approach (SM-SRC) only SRC are included in the
correlation function and the final state of the residual nucleus is a pure 
two-hole state in the target. For instance, the ground state of $^{14}$C is a 
($p_{1/2}$)$^{-2}$ hole in $^{16}$O.

In the more sophisticated approaches \cite{Giu97} and \cite{barb}, the 
TOF is obtained from the the first calculations of the two-nucleon spectral 
function of $^{16}$O, where SRC, TC and LRC are included with some
approximations but consistently. In both calculations the TOF's for low-lying
states of the residual nucleus are obtained partitioning the Hilbert space onto
two subspaces where SRC and LRC are separately calculated. LRC and the
long-range part of TC are calculated using the self-consistent Green's 
function formalism \cite{BD} in an appropriate harmonic-oscillator (h.o.) 
basis, large enough to account for the main collective features that influence 
the pair removal amplitudes. The effects of SRC due to the central and tensor 
part at high momenta are included by defect functions, which are solutions of a  
Bethe-Goldstone equation where the Pauli operator considers only 
configurations outside the model space where LRC are calculated. Different 
defect functions are obtained for different relative states and the Bonn C
NN-potential \cite{MaH87} is used in the calculations.
The TOF is given by a combination of components of relative and CM motion, 
that are different for different final states, where the coefficients are the 
two-nucleon removal amplitudes which include LRC, and the defect functions, 
which contain SRC, are intertwined in a complicated manner. 

In the more recent calculation \cite{barb} some improvements have been
included with respect to the previous work \cite{Giu97}: i) the 
non-locality of the Pauli operator is computed exactly, resulting in a larger 
number of defect functions with a more complicate state dependence; ii) 
the  evaluation of nuclear structure effects related to the fragmentation of 
the s.p. strength has been improved by applying a Faddeev technique to 
the description of the internal propagators in the nucleon self-energy. 
Moreover, both pp- and pn-pairs are calculated in \cite{barb}, while only the 
pp-case is considered in \cite{Giu97}. The numerical results presented in this
contribution are obtained with the more recent TOF's from \cite{barb} (SF-B).

In \cite{Giu99} the defect 
functions for pn-knockout off $^{16}$O  are calculated within the framework of 
the coupled cluster (CC) method, using the so-called 
$S_2$ approximation, and employing the Argonne V14 NN-potential \cite{v14}.
Also in this calculation (SF-CC) the defect functions include SRC and TC and 
have a complicate state dependence. LRC, however, are accounted for in the TOF 
in a simpler way and only knockout of nucleons from the $0p$  
shell is considered.  

An alternative procedure is proposed in \cite{Kadrev,TDM}, where the TOF 
is obtained from the asympotic behavior of the two-body density matrix (TDM).
This procedure avoids the calculation of the spectral function, but depends 
strongly and relies on the availability of reliable calculations of the TDM. 
The applicability of the procedure is shown in \cite{Kadrev}, where the TOF
obtained from a simple TDM, which is calculated within the Jastrow correlation 
method \cite{DKA} (JCM) and incorporates only SRC, is able to give 
numerical results that under many aspects are qualitatively similar to the 
results produced by more sophisticated approaches. The TDM of \cite{DKA} is 
at present the only TDM available for these calculations. It would be 
interesting to check the procedure with a more complete TDM. 

\subsection{Center-of-Mass Effects and Orthogonality}
\label{sec:CM}

In the calculation of the transition amplitude of Eq.~(\ref{eq:jq}), for the 
three-body system consisting of the two nucleons, 1 and 2, and of the residual 
nucleus $B$, it is natural to work with CM coordinates \cite{GP,cm1,cm2,jack} 
\begin{equation}
{\mbox{\boldmath $r$}}_{1B} = {\mbox{\boldmath $r$}}_{1} - 
{\mbox{\boldmath $r$}}_{B}, \,\,\,\,\, 
{\mbox{\boldmath $r$}}_{2B} = {\mbox{\boldmath $r$}}_{2} - 
{\mbox{\boldmath $r$}}_{B},\,\,\,\,\,
{\mbox{\boldmath $r$}}_{B} = \sum _{i=3} ^{A} {\mbox{\boldmath $r$}}_{i}/ (A-2).
\label{eq:cm}
\end{equation}
The conjugated momenta are given by
\begin{eqnarray}
{\mbox{\boldmath $p$}}_{1B} &=& \frac{A-1}{A}
{\mbox{\boldmath $p'$}}_{1} - \frac{1}{A} {\mbox{\boldmath $p'$}}_{2}
- \frac{1}{A} {\mbox{\boldmath $p$}}_{B} \,\, , \nonumber \\ 
{\mbox{\boldmath $p$}}_{2B} &=& -\frac{1}{A}
{\mbox{\boldmath $p'$}}_{1} + \frac{A-1}{A} {\mbox{\boldmath $p'$}}_{2}
- \frac{1}{A} {\mbox{\boldmath $p$}}_{B} \,\, ,  \nonumber \\
{\mbox{\boldmath $P$}} &=& 
{\mbox{\boldmath $p'$}}_{1} + {\mbox{\boldmath $p'$}}_{2}
+  {\mbox{\boldmath $p$}}_{B} \,\, , 
\label{eq:mom}
\end{eqnarray}
where  $\p_{B}=\q-\p'_1-\p'_2$ is the momentum of the residual nucleus in the
laboratory frame.

With the help of these relations, the transition amplitude of 
Eq.~(\ref{eq:jq}) is obtained in the following form \cite{cm1,cm2}
\begin{equation}
 J^{\mu}({\mbox{\boldmath $q$}})  = \int
{ \psi}_{\rm{f}}^{*}({\mbox{\boldmath $r$}}_{1B},
{\mbox{\boldmath $r$}}_{2B})
V^{\mu}({\mbox{\boldmath $r$}}_{1B},{\mbox{\boldmath $r$}}_{2B}) 
{ \psi}_{\rm{i}}
({\mbox{\boldmath $r$}}_{1B},{\mbox{\boldmath $r$}}_{2B})
{\rm d}{\mbox{\boldmath $r$}}_{1B} {\rm d}{\mbox{\boldmath $r$}}_{2B},  
 \label{eq:jqcm}
\end{equation}
where for the OB current $j^{(1)\mu}$ 
\begin{equation}
V^{\mu}({\mbox{\boldmath $r$}}_{1B},{\mbox{\boldmath $r$}}_{2B}) =
\exp \left( i {\mbox{\boldmath $q$}} \frac {A-1}{A}{\mbox{\boldmath $r$}}_{1B}
\right) \exp \left(-i {\mbox{\boldmath $q$}} \frac {1}{A}
{\mbox{\boldmath $r$}}_{2B}\right) 
j^{(1)\mu}({\mbox{\boldmath $r$}_{1B}})  \,\,+ \,\, \left( 1 \leftrightarrow 2 \right).
\label{eq:opcm}
\end{equation}
Similar expressions are obtained for the TB currents \cite{cm1,cm2}.

In spite of the fact that an OB operator cannot affect two particles if they are 
not correlated, it can be seen from Eqs.~(\ref{eq:jqcm}) and (\ref{eq:opcm}) 
that in the CM frame the transition operator becomes a TB operator even in the 
case of an OB nuclear current. 
Only in the limit $A \rightarrow \infty$, CM effects vanish and the expression 
in Eq.~(\ref{eq:jq}) becomes zero for a pure OB current when the matrix
element is calculated using orthogonalized s.p. wave functions.
This means that, due to this CM effect, for finite nuclei the OB current can 
give a contribution to the cross section of two-particle emission even without 
correlations in the nuclear wave functions \cite{cm1,cm2}.

Independently of the specific prescriptions adopted in the calculations, a 
conceptual problem arises in the model, where the 
initial and final states, $\psi_{\rm{i}}$ and $\psi_{\rm{f}}$,
which are eigenfunctions of an energy-dependent optical-model Hamiltonian, are, 
as such, not orthogonal. 
Indeed, the process involves transitions between bound and continuum states 
which must be orthogonal, since they are eigenfunctions of the full nuclear 
many-body Hamiltionian. Orthogonality is in general lost in a model when the 
description is restricted to a subspace where other channels are suppressed.
The description of direct knockout reactions in terms of the eigenfunctions of 
a complex energy-dependent optical potential considers only partially the 
contribution of competing inelastic channels. 
A consequence of the lack of orthogonality between the initial and the final 
state is that Eq.~(\ref{eq:jqcm}) may contain a spurious contribution, since it 
does not vanish when the transition operator  $V^\mu$  is set equal to 1.
The use of an effective nuclear current operator would remove the orthogonality 
defect besides taking into account space truncation effects \cite{Oxford,ort}.
In the usual approach, however, the replacement of the effective operator 
by the bare nuclear current operator may introduce a spurious contribution.  

In our earlier work \cite{GP97,Giu97,barb,Giu98,Giu99,Kadrev} the  
spuriosity was removed subtracting from the transition amplitude 
the contribution of the OB current without correlations in the nuclear wave 
functions. This prescription, however, subtracts, together with the spuriosity,
also the CM effect given by the TB operator in Eq.~(\ref{eq:opcm}), that is 
present in the OB current independently of correlations in the nuclear wave
function, and that is not spurious.
Moreover, the definition of a nuclear wave function without correlations is
clear in the simpler approach SM-SRC, where the uncorrelated SM pair function 
is multiplied by a correlation function, but it is less clear in a more
sophisticated approach, like SF-B, where different types of correlations are
intertwined in the TOF. In this case the prescription is applied subtracting 
the part of the TOF without the defect functions. The defect functions, however,
include only SRC, while different correlations are intertwined in the remaining 
part of the TOF that is subtracted.
Therefore, with this prescription also part of the contribution of 
correlations is subtracted. 

A more accurate procedure to get rid of the spuriosity is adopted in 
\cite{cm1,cm2}, where orthogonality between the initial and final states is
enforced by means of a Gram-Schmidt orthogonalization \cite{ortho}. In this 
approach each one of the two s.p. distorted scattering wave functions is 
orthogonalized to all the s.p. SM wave functions that are used to calculate 
the TOF, {\it i.e.}, for the TOF of \cite{barb}, to the h.o. states of the 
basis used in the calculation of the spectral function, which range from  
the $0s$ up to the $1p$-$0f$ shell, or, for the simpler SM-SRC approach, to 
the s.p. states used to calculate the SM pair function. 

The procedure to enforce orthogonality between s.p. bound and scattering states
is done accordingly with the definition of the spurious contribution, that in 
the limit $A \rightarrow \infty$, where CM effects are neglected, 
the transition amplitude vanishes for a pure OB current sandwiched between 
orthogonalized s.p. wave functions.

The orthogonalization procedure adopted in our recent work \cite{cm1,cm2} 
removes the spurious contribution and includes all the CM and correlation 
effects that were subtracted with the previous prescription and that are not 
spurious.
The relevance of these effects depends on the type 
of reaction, on the kinematics, and on the treatment 
of the theoretical ingredients of the model \cite{cm1,cm2}. In many situations 
the differences between the results obtained with the two procedures are small. 
There are, however, situations where these differences are large, and they are 
very large in the super-parallel kinematics. This is a kinematical setting of 
particular interest, since the recent  $^{16}$O(e,e$'$pp)$^{14}$C  \cite{Ros00} 
and  $^{16}$O(e,e$'$pn)$^{14}$N  \cite{Duncan} experiments carried out at MAMI 
are both centred on the same super-parallel kinematics. 

The super-parallel kinematics was originally proposed \cite{GP} 
since it is particularly favourable to emphasize short-range effects and has 
been widely investigated in our work. Recent results
\cite{barb,cm1,cm2,delta,sch1,sch2} have shown that this kinematic setting is 
indeed very sensitive to correlations, but it is also very sensitive to all 
the other ingredients of the model. Such a great sensitivity makes the
super-parallel kinematics very interesting although not very suitable to 
isolate the specific contribution of correlations from other competing 
contributions.    
The achievement of this goal requires experiments in a wider range of kinematics 
which mutually supplement each other.
These arguments are illustrated in the next section with some numerical 
examples.

\section{Results}
\label{sec.results}

In the super-parallel kinematics the two nucleons are ejected parallel and 
anti-parallel to $\q$ and, for a fixed value of $q$ and of the energy 
transfer $\omega$,  it is possible to explore, 
for different values of the kinetic energies of the outgoing nucleons, all 
possible values of the recoil-momentum $p_{\mathrm{B}}$. 

\begin{figure}[ht]
\begin{center}
\includegraphics[height=120mm,width=100mm]{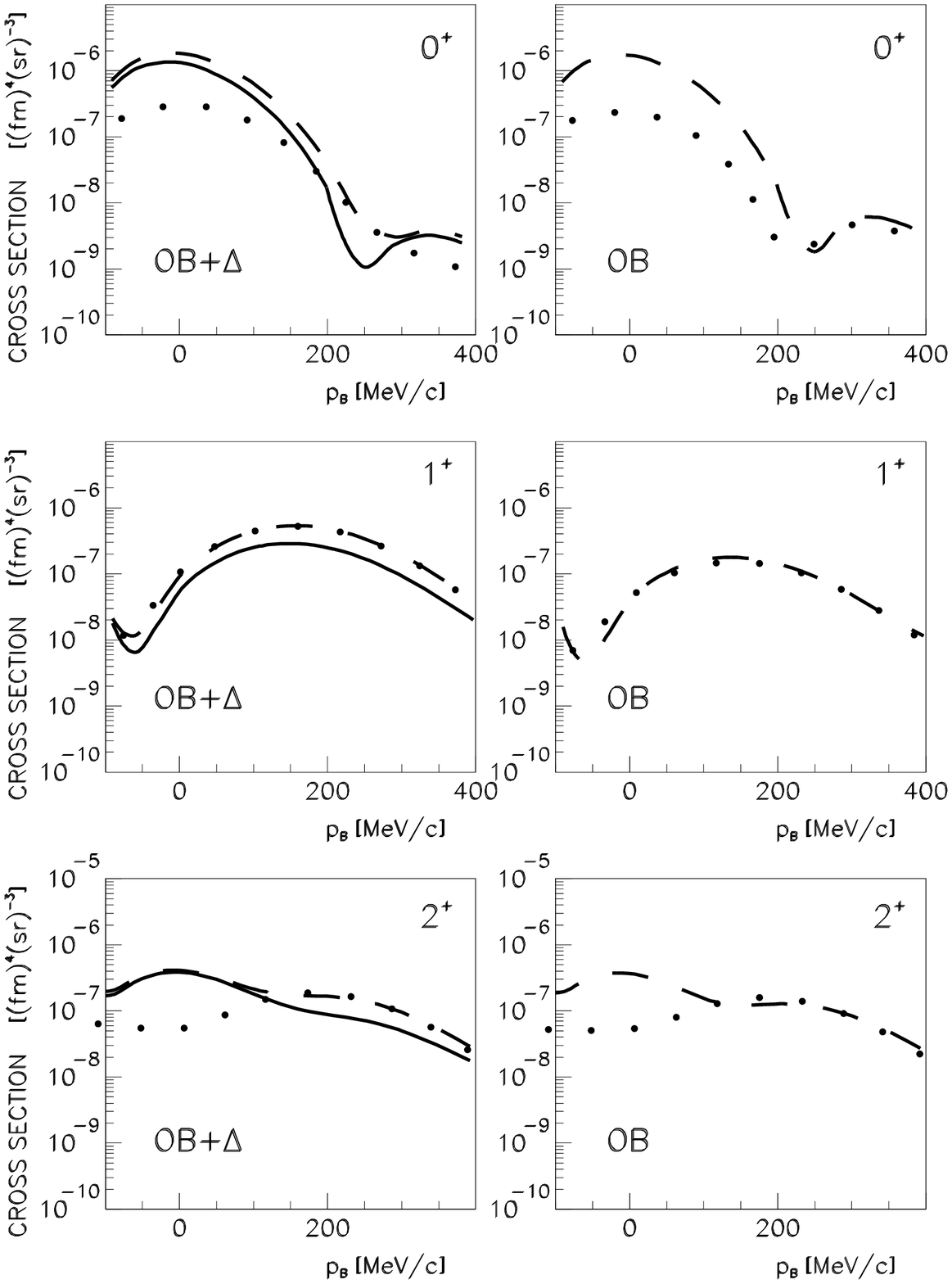}
\vspace{2mm}\caption[]{The differential cross section of the reaction 
$^{16}$O(e,e$'$pp)$^{14}$C to the low-lying states of $^{14}$C: the $0^+$ ground 
state, the  $1^+$ state at 11.31 MeV, and the  $2^+$ state at 7.67 MeV. 
The super-parallel kinematics of the MAMI experiment \cite{Ros00} is 
considered, with an incident electron energy $E_0=855$ MeV, $\omega=215$ MeV 
and $q=316$ MeV/$c$. 
Different values of $\p_{\mathrm{B}}$ are obtained changing the kinetic 
energies of the two outoging nucleons. Positive (negative) 
values of $p_{\mathrm{B}}$ refer to situations where $\p_{\mathrm{B}}$  is 
parallel (anti-parallel) to $\q$. The final results given by sum of the OB and 
TB $\Delta$-current are displayed in the left panels, the separate contribution 
of the OB current is shown in the right panels. The TOF from \cite{barb} (SF-B) 
and the DW approximation for FSI are used in the calculations. 
The dotted lines give the results of \cite{barb}, the dashed and solid lines 
are obtained in the orthogonalized approach with different parametrizations of 
the TB currents, {\it i.e.} the previous unregularized pararametrization, as 
in \cite{barb}, (dashed) and the Bonn parametrization (solid).
\label{fig:fig1}
}
\end{center}
\end{figure}

In Fig.~\ref{fig:fig1} the differential cross sections of the 
$^{16}$O(e,e$'$pp)$^{14}$C reaction are displayed, as a function of 
$p_{\mathrm{B}}$, for transitions to different states of $^{14}$C:
the $0^+$ ground state, the  $1^+$ state 
at 11.31 MeV, and the  $2^+$ state at 7.67 MeV.  The calculations have been 
performed in the super-parallel kinematics of the MAMI experiment \cite{Ros00}, 
with the SF-B overlap function and the DW approximation for FSI. 
Different shapes are obtained for the cross sections to the three final states. 
The shape of the recoil-momentum distribution is 
basically driven by the CM orbital angular momentum $L$ of the knocked out 
pair. Different partial waves of relative and CM motion contribute to the TOF. 
Each transition is characterized by different components, with specific values 
of $L$. The relative weights of these components, which are given by the 
two-nucleon removal amplitudes included in the TOF, determine the shape of the 
momentum distribution \cite{barb}.  

The results of the orthogonalized approach and of the previous 
prescription to remove the spuriosity are compared in the figure.
The difference depends on the final state.
For the $0^+$ ground state the orthogonalized approach gives a large enhancement
of the cross section calculated with the OB current. The result is shown in 
the right panel, where it can be seen that the enhancement is 
large for values of $p_{\mathrm{B}}$ up to about 300 MeV/$c$, and is a factor of 
about 5 in the maximum region. It is shown in \cite{cm1} that the enhancement 
is mostly due to the CM effects included in the new calculations and not to 
the different treatment of the spuriosity or to the restoration  of 
orthogonality between the initial and final-state wave functions.  
The difference between the results of the two approaches is only slightly 
reduced in the final cross sections, shown in the left panel, where also the  
$\Delta$-current is included. 
The cross section is dominated by the OB current and only a minor role is 
played by the $\Delta$-current. As a consequence, also the difference between 
the results obtained in the orthogonalized approach with the Bonn and the 
unregularized parametrizations, that are also shown in the
figure, is small. A small reduction of the cross section is given by the 
Bonn parametrization.

For the $1^+$ state the difference between the results of the orthogonalized 
approach and the previous 
results is practically negligible. In this case the $\Delta$-current gives a 
significant enhancement of the OB cross section, which is larger with the 
unregularized than with the Bonn parametrization. 
For the $2^+$ state the orthogonalized approach gives a significant enhancement 
of the cross section at low values of $p_{\mathrm{B}}$ and a negligible effect 
at higher momenta. The main contribution to the cross section is given by
the OB current, the two parametrizations of the $\Delta$-current 
give a negligible difference at low values of $p_{\mathrm{B}}$ and a small 
reduction at larger momenta is obtained with the Bonn parametrization. 
 
\begin{figure}[ht]
\begin{center}
\includegraphics[height=55mm,width=100mm]{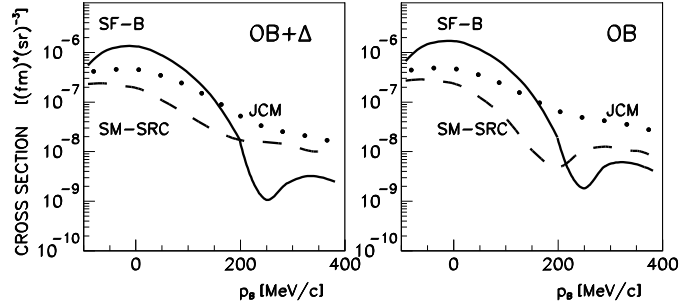}
\vspace{2mm}\caption[]{The differential cross section of the reaction 
$^{16}$O(e,e$'$pp)$^{14}$C to the $0^+$ ground state in the same super-parallel
kinematics as in Fig.~1. The final results given by sum of the OB and 
TB $\Delta$-current (calculated with the BONN parametrization) are displayed 
in the left panel, the separate contribution of the OB current is shown in the
right panel. The curves are obtained in the orthogonalized approach with 
different TOF's: SF-B (solid), SM-SRC (dashed), JCM (dotted). The DW 
approximation is used for FSI.
\label{fig:fig2}
}
\end{center}
\end{figure}

The cross sections are very sensitive to correlations and to their
treatment in the TOF. An example is shown in Fig.~\ref{fig:fig2}, for the 
reaction $^{16}$O(e,e$'$pp)$^{14}$C to the $0^+$ ground state in the same 
super-parallel kinematics as in  Fig.~\ref{fig:fig1}. The cross sections 
calculated in the orthogonalized approach with three different TOF's are 
compared: the TOF from the spectral function (SF-B), from the
simpler approach (SM-SRC), and from the asymptotic behaviour of a TDM  
including only Jastrow correlations (JCM). The three cross sections are very 
different, both in magnitude and shape, and  all of them are
dominated by the OB current, at least at lower values of $p_{\mathrm{B}}$. 

\begin{figure}[ht]
\begin{center}
\includegraphics[height=55mm,width=100mm]{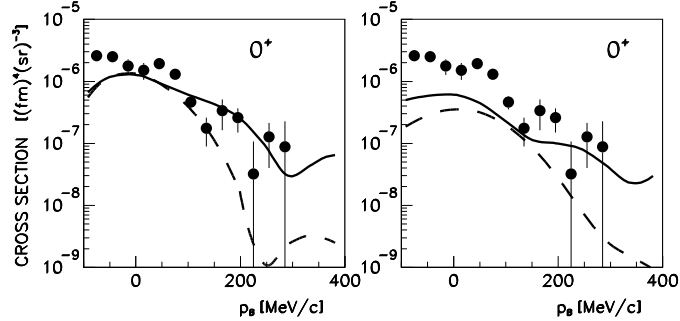}
\vspace{2mm}\caption[]{The differential cross section of the reaction 
$^{16}$O(e,e$'$pp)$^{14}$C to the $0^+$ ground state in the same super-parallel 
kinematics as in Fig.~1. Solid and dashed lines are obtained with the DW-NN and 
DW approaches for FSI. The results of the orthogonalized approach and of the 
previous prescription to remove the spuriosity  are displayed in the left and 
right panels, respectively. The TOF from SF-B and the Bonn parametrization 
for the $\Delta$-current are used in the calculations. Experimental data 
from \cite{Ros00}. 
\label{fig:fig3}
}
\end{center}
\end{figure}

The effect of the mutual interaction between the two outgoing protons (NN-FSI),
which has been neglected in the calculations presented till now, is of
particular relevance for the (e,e$'$pp) reaction in the super-parallel 
kinematics \cite{sch1,sch2}. An example is shown in  
Fig.~\ref{fig:fig3}, where the cross sections calculated in the DW-NN and DW
approaches are compared for the reaction $^{16}$O(e,e$'$pp)$^{14}$C to the $0^+$
ground state. Calculations have been performed with the 
TOF from SF-B and the Bonn parametrization for the $\Delta$-current. The
comparison with the experimental data \cite{Ros00} is also shown in the figure.

The contribution of NN-FSI is very large, both in the orthogonalized approach 
and with our previous prescription, and produces a strong enhancement of the 
cross section that is particularly large just when the cross section calculated in the DW approach 
is small. The contribution of NN-FSI is very large at large values of 
$p_{\mathrm{B}}$.  The DW-NN cross sections calculated in the orthogonalized and
in the previous approaches are very different. The difference is mainly due to
the CM and correlation effects that were neglected in the previous approach. 
NN-FSI can be considered an effect of correlations between the two nucleons 
in the final state. 

The comparison with the MAMI data \cite{Ros00} in Fig.~\ref{fig:fig3} 
emphasizes the important contribution of the CM and correlation effects 
included in the new orthogonalized calculations, the essential 
role played by NN-FSI, as well as the need of a careful 
treatment of different types of correlations, which are consistently included 
in the TOF from SF-B. A different choice of the theoretical ingredients in 
Figs.~\ref{fig:fig1}-\ref{fig:fig3} can give huge differences. 
The most refined version of the model, with the best available ingredients 
(the DW-NN approach with enforced orthogonality between initial and final
states, the TOF from SF-B, the Bonn parametrization for the TB currents) 
is able to give in the best agreement in comparison to data 
and a reasonable description of the $^{16}$O(e,e$'$pp) data.

\begin{figure}[ht]
\begin{center}
\includegraphics[height=55mm,width=100mm]{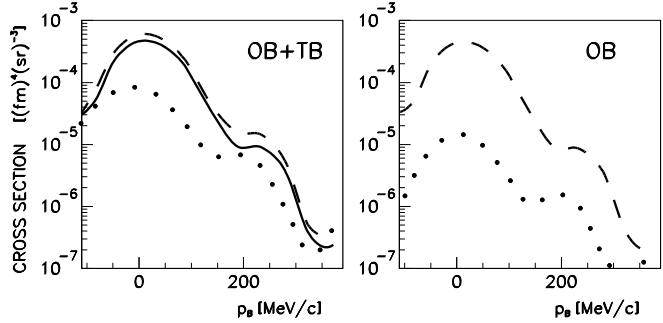}
\vspace{2mm}\caption[]{The differential cross section of the reaction 
$^{16}$O(e,e$'$pn)$^{14}$N to the $1_2^+$ state at 3.95 MeV in the same
super-parallel kinematics as in Fig.~1. The proton is emitted parallel and the 
neutron antiparallel to the momentum transfer.
The final results given by sum of the OB and TB currents are displayed in the 
left panel, the separate contribution of the OB current is shown in the right 
panel. Line convention as in Fig.~1.
\label{fig:fig4}
}
\end{center}
\end{figure}

In Fig.~\ref{fig:fig4} the differential cross section of the reaction 
$^{16}$O(e,e$'$pn)$^{14}$N is displayed for the transition to the $1_2^+$ 
excited state of $^{14}$N at 3.95 MeV. This is the state that is mostly 
populated in pn-knockout \cite{Duncan,pn}. Results for this transition are 
therefore of particular interest. Calculations have been performed, with the 
TOF from SF-B and the DW approximation for FSI,
in the same super-parallel kinematics already considered in 
Figs.~\ref{fig:fig1}-\ref{fig:fig3} and realized in the 
$^{16}$O(e,e$'$pn)$^{14}$N  experiment at MAMI \cite{Duncan}. 
The previous results of \cite{barb} are compared in the figure with the results 
of the orthogonalized approach. 

The CM effects included in the orthogonalized approach give a huge 
enhancement of the cross section calculated with the OB current. The result 
is shown in the right panel of Fig.~\ref{fig:fig4}.  The enhancement is larger 
than one order of magnitude at low recoil momenta and is only slightly reduced 
at higher momenta. The difference between the cross sections of the two
approaches is reduced when the TB 
currents are added. The result is shown in the left 
panel. In the previos calculations \cite{barb} the cross section is 
dominated by the TB currents, in particular by the $\Delta$-current. In
contrast, the OB current dominates the cross section in the orthogonalized 
approach and only a small enhancement is given by the TB currents. 
The differences obtained in the orthogonalized approach with the Bonn and the 
unregularized  parametrizations for the TB currents are appreciable, although 
not very important. The Bonn parametrization reduces the cross section by at 
most 30-40 \%.  
In the final cross section the difference between the results of the two
approaches remains large, a bit less than one order of magnitude at low 
momenta, in the maximum region, and it is still sizable, although considerably 
reduced, at higher momenta. 

\begin{figure}[ht]
\begin{center}
\includegraphics[height=55mm,width=100mm]{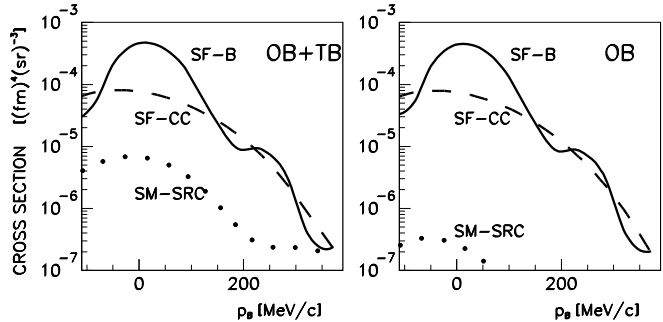}
\vspace{2mm}\caption[]{The differential cross section of the reaction 
$^{16}$O(e,e$'$pn)$^{14}$N to the $1_2^+$ state at 3.95 MeV in the same
super-parallel kinematics as in Fig.~4. The curves  are obtained with different 
TOF's: SF-B (solid), SF-CC (dashed), SM-SRC (dotted). The final results given 
by sum of the OB and TB currents are displayed in the left panel, the separate 
contribution of the OB current is shown in the right panel. Calculations are 
performed in the orthogonalized approach, with the DW approximation for FSI, 
and the Bonn parametrization for the TB currents.
\label{fig:fig5}
}
\end{center}
\end{figure}

 The sensitivity to  the treatment of correlations is shown in 
 Fig.~\ref{fig:fig5}, where the cross sections of the reaction 
$^{16}$O(e,e$'$pn)$^{14}$N  to the $1_2^+$ state calculated in the
super-parallel kinematics with the TOF's from SF-B, SF-CC, and  the simpler 
SM-SRC are compared.  
The calculations have been performed with the orthogonalized approach, the DW 
approximation for FSI, and the Bonn parametrization for the TB currents. 
The three TOF's give large differences, both in the shape and in the magnitude 
of the cross section.
With the simpler SM-SRC, where only SRC are taken into account, the 
contribution of the OB current is negligible and is up to about three orders 
of magnitude lower than the one obtained with SF-B. 
When the TB currents are added, the SM-SRC cross section is enhanced by about 
one order of magnitude, the difference between the SM-SRC and SF-B cross 
sections is reduced but remains very large, up to about two orders of 
magnitude in the maximum region and somewhat smaller at high values of 
$p_{\mathrm{B}}$. 

The SM-SRC cross section is dominated by the TB currents and, as such, it is
practically unaffected by the use of orthogonalized s.p. bound and scattering
wave functions. Incidentally, we note that for the separate contribution of 
the OB current the CM effects included in the orthogonalized approach 
give with SM-SRC a much smaller effect than with SF-B and SF-CC, whose cross
sections are, in both cases, dominated by the OB current. The dominance of the
OB current is due to the very large enhancement of the OB contribution 
produced, both with SF-B and SF-CC, by the CM effects included the 
orthogonalized approach. In the previous calculations with SF-B \cite{barb} 
and SF-CC \cite{Giu99} the cross section to the $1_2^+$ state was in both 
cases dominated by the TB $\Delta$-current.

The much larger contribution of the OB current found with SF-B and SF-CC 
emphasizes the crucial role played by TC, that are very important in 
pn-emission and  are neglected in the simpler SM-SRC calculation. In the SF-B 
and SF-CC overlap functions, SRC and TC are accounted for consistently in the 
defect functions, which are calculated in the two TOF's within different 
methods. 

In  Fig.~\ref{fig:fig5} the SF-B cross section is generally larger than the 
SF-CC one.  The SF-B result overshoots the SF-CC one up to a factor of 6 in the 
maximum region. The differences are strongly reduced for values of
$p_B$ greater than 100 MeV/$c$. The differences between the two cross sections 
are due to the different treatment of SRC, TC and LRC. 
A more complete calculation of LRC in an extended shell-model basis is 
performed in SF-B \cite{barb}, where the normalization of the two-nucleon 
overlap amplitudes is higher than in SF-CC. The difference in the 
shape of the cross section is due to the different mixing of configurations in 
the two calculations. 

The results in Fig.~\ref{fig:fig5}  emphasize the need of a careful and
consistent treatment of different correlations in the TOF.  
The contribution of NN-FSI, that is very large in the super-parallel
kinematics for the (e,e$'$pp) reaction, is strongly reduced in the same 
kinematics for pn-knockout, and it is quite moderate for the  $1_2^+$ 
state \cite{cm2}. 

The strong enhancement of the cross section to the $1_2^+$ state at low values
of $p_{\mathrm{B}}$, that is due to the CM effects included in the 
orthogonalized approach, is able to resolve the discrepancies found 
\cite{Duncan} in comparison with the experimental data  and give a much 
better agreement \cite{secondpn}. A careful comparison with the data will be 
presented in a forthcoming paper \cite{secondpn}.

\begin{figure}[ht]
\begin{center}
\includegraphics[height=100mm,width=100mm]{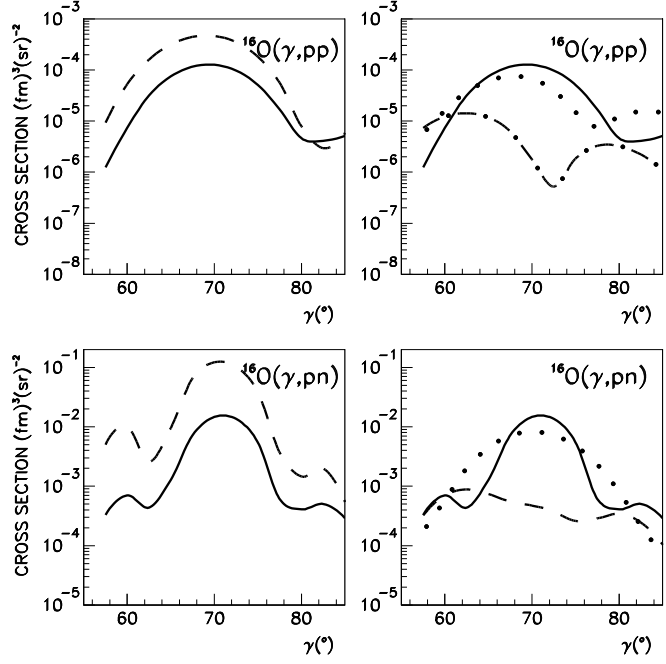}
\vspace{2mm}\caption[]{The differential cross section of the reactions 
$^{16}$O($\gamma$,pp)$^{14}$C$_{\mathrm{g.s.}}$ (top panels) and  
$^{16}$O($\gamma$,pn)$^{14}$N to the $1_2^+$ state (bottom panels), in a 
coplanar symmetrical kinematics at $E_\gamma=400$ MeV, as a function of the 
scattering angle $\gamma$ of the outgoing nucleons. The results obtained with 
the TOF from SF-B for the Bonn (solid) and the previous unregularized 
pararametrization (dashed) for the TB currents are compared in the left panels. 
The results obtained with the Bonn parametrization for different TOF's are 
compared in the right panels: SF-B (solid), SM-SRC (dotted), JCM (dot-dashed), 
SF-CC (dashed). 
The calculations are performed in the orthogonalized approach and with the DW
approximation for FSI. 
\label{fig:fig6}
}
\end{center}
\end{figure}

The super-parallel kinematics is particularly sensitive to all the 
ingredients of the model. Different results can be obtained in
different situations. A suitable choice of kinematics can be helpful 
to disentangle specific contributions and reduce the 
uncertainties on the theoretical ingredients.
An example of a different kinematics is shown  in Fig.~\ref{fig:fig6}, 
where the cross sections of the reactions 
$^{16}$O($\gamma$,pp)$^{14}$C$_{\mathrm{g.s.}}$ and 
$^{16}$O($\gamma$,pn)$^{14}$N to the $1^+_2$ state are displayed, for a coplanar 
symmetrical kinematics with an incident photon energy $E_\gamma=400$ MeV. 
In symmetrical kinematics the two
nucleons are ejected at equal energies and equal but opposite angles with
respect to $\q$, and different values of $p_{\mathrm{B}}$ are obtained 
changing the scattering angles of the two outgoing nucleons. In this kinematics 
the difference between the results of the orthogonalized and the 
previous prescription are generally small \cite{cm1,cm2}, and also the 
effects of NN-FSI are very small. The cross sections are, however, very
sensitive to the parametrization of the TB-currents. The results 
of the Bonn and the unregularized parametrizations, which are compared
in the left panel of the figure, differ by about one order of magnitude.
Also for this kinematics the results are very sensitive to the treatment
of correlations in the TOF. The cross sections calculated with different TOF's, 
which are compared in the right panels, exhibit very large 
differences, both in magnitude and shape.

\section{Conclusions}
\label{sec.conc}

Recent improvements in the theortical description of electromagnetic two-nucleon
knockout have been reviewed: the sensitivity to the correlations included  
in the nuclear wave functions, the parametrization used for the TB currents, 
the effects of final-state interactions, and the role of center-of-mass effects 
in connection with the problem of the lack of orthogonality between  
bound and scattering states obtained by the use of an 
energy-dependent optical potential. Numerical examples have been presented for
 pp- and pn-knockout off  $^{16}$O also in comparison with the available data. 

The most refined version of the model, with the best available ingredients, is
able to give a reasonable, although not perfect, description of the experimental
data. This result is an indication that the main theoretical
ingredients contributing to the cross section seem reasonably under control.
The cross sections are very sensitive to correlations and to their
treatment. Different type of correlations, of short-range and long-range type, 
are important and require a careful and consistent treatment. An essential role
is played by tensor correlations in pn-knockout.
The complexity of the necessary model ingredients makes it  difficult to 
extract clear and unambigous information on correlations from one or 
two ``ideal'' kinematics. 
If we want to isolate the contribution of correlations, data are needed in 
various kinematics which mutually supplement each other. 
Close collaboration between theorists and experimentalists is necessary to 
achieve this goal.

\section*{}

\end{document}